# Direct evidence of interfacial coherent electron-phonon coupling in single-unit-cell FeSe film on Nb-doped SrTiO$_3$

Mengdi Zhang[1,2], Xiaotong Jiao[2], Mingxia Shi[2], Wenfeng Dong[2], Cui Ding[1,2], Yubin Wang[2], Tingxiao Qin[1], Haiyun Liu[1*], Lili Wang[2,3*], Zhenyu Zhang[5], Qi-Kun Xue[1,2,3,6], Qihua Xiong[1,2,3,4*]

[1] Beijing Academy of Quantum Information Sciences, Beijing 100193, P.R. China

[2] State Key Laboratory of Low-Dimensional Quantum Physics and Department of Physics, Tsinghua University, Beijing 100084, P.R. China

[3] Frontier Science Center for Quantum Information, Beijing 100084, P.R. China

[4] Collaborative Innovation Center of Quantum Matter, Beijing 100084, P.R. China

[5] International Center for Quantum Design of Functional Materials (ICQD), Hefei National Laboratory for Physical Sciences at Microscale, and Synergetic Innovation Center of Quantum Information and Quantum Physics, University of Science and Technology of China, Hefei, Anhui 230026, P.R. China

[6] Southern University of Science and Technology, Shenzhen 518055, China

The interface-enhanced superconductivity in monolayer iron selenide (FeSe) films on SrTiO$_3$ has been actively pursued in the past decade. Although a synergistic effect between interfacial charge transfer and interfacial electron-phonon coupling (EPC) is proposed to be responsible for the mechanism, the microscopic nature of the interfacial EPC in the enhancement of superconductivity remains highly controversial. Herein we experimentally reveal that a coherent optical phonon mode at 4.2 THz from the SrTiO$_3$ substrate couples to FeSe electrons and modulates the quasiparticle relaxations using ultrafast pump-probe spectroscopy. This mode originates from the antiferrodistortive (AFD) transition in SrTiO$_3$ and is significantly stronger in the presence of monolayer FeSe than that in purely FeTe-capped Nb-doped SrTiO$_3$. Pump fluence and temperature-dependent spectroscopy measurements suggest that SrTiO$_3$ substrate facilitates the stabilization of FeSe structure and possibly prevents the occurrence of nematic phase transition, supporting that SrTiO$_3$ substrate modifies the electronic structure of monolayer FeSe through a strong interfacial EPC strength as large as 0.77. Our results provide unprecedented direct evidence that the strong coupling of SrTiO$_3$ coherent phonon to FeSe electrons is indeed responsible for the high-temperature superconductivity in monolayer FeSe and SrTiO$_3$ heterostructure.

**Introduction**

Over the past decade, the discovery of interfacial superconductivity in single-unit-cell (1 UC) FeSe films grown on SrTiO$_3$ (001) (STO) substrate[1-8] has sparked intense research interest, with a remarkably high critical temperature ($T_c$) of 65−109 K reported, which sets a record high $T_c$ for iron-based superconductors and challenges the conventional understanding of superconducting pairing mechanisms. The origin of the interfacial $T_c$ enhancement in this system remains an open issue that has attracted numerous theoretical and experimental investigations. Two contributing factors, the interfacial charge transfer and electron-phonon coupling (EPC), are proposed to cooperatively boost the superconductivity. The electronic structure of monolayer FeSe resembles that of heavily electron-doped A$_x$Fe$_{2-y}$Se$_2$ (A = K, Cs and so on, $T_c$ > 30 K)[9,10] and Li$_{1-x}$Fe$_x$OHFeSe ($T_c$ ~ 41 K)[11,12], with four electron pockets around the Brillouin zone (BZ) corners but the absence of a hole pocket at the BZ center. Previous experiments have examined the electron doping effect by depositing potassium into 3-50 UC FeSe films on various substrates[13-15] or ionic liquid-gating onto the FeSe films or flakes[16-18], showing promoted $T_c$ up to 48 K, nonetheless significantly lower than that of 1 UC FeSe/STO. These demonstrate that electron doping alone is unlikely to fully induce the $T_c$ enhancement in 1 UC FeSe/STO.

Previous experimental observations provided evidence of interfacial EPC as a primary cause of the extra $T_c$ enhancement. Angle-resolved photoemission spectroscopy (ARPES) experiments have disclosed the "replica band"[4] at the energy separation of ~ 100 meV featured by O$^{16}$ ↔ O$^{18}$ isotope dependences in 1 UC FeSe/STO firstly[19] and subsequently similar "replica bands" in the FeSe films grown on BaTiO$_3$ (001)[20], STO (110)[21], and rutile TiO$_2$ (100)[22], pinning down the existence of interfacial EPC. High-resolution electron energy loss spectroscopy (HREELS) experiments demonstrated that the Fuch-Kliewer (F-K) modes of STO at the consistent energy of ~ 97 meV can penetrate into FeSe films[23,24], leading to the formation of the dynamic interfacial polarons[25]. In addition, multiple experimental investigations provided clues of correlation between interfacial EPCs and superconductivity , including both mutual inductance and Raman spectra anomalies corresponding to a broadened ferroelectric transition near the gap-opening temperature ~50 K[26], a large EPC constant λ = 0.48 by ultrafast dynamics[27], the dip-hump features of boson modes by scanning tunneling spectroscopy (STS)[28,29], and the poping-up vibrational modes of the surface Ti-O bonding by ultraviolet photoemission spectroscopy (UPS) associated with gap opening[30]. However, direct conclusive evidence for the enhanced EPC as well as its correlation with superconductivity is still in debate. On the other hand, the effect of EPC on superconductivity is still theoretically controversial[31,32], which requires more direct evidence from quasiparticle dynamics.

Ultrafast pump-probe spectroscopy allows unveiling quasiparticle dynamics in time domain from femtosecond to picosecond, such as electron-electron interactions, EPC and spin mechanics[27,33-36]. In this work, we report on a coupling enhancement between an STO coherent optical phonon and FeSe electrons in the FeSe/STO heterostructure using transient reflectivity measurements. Photoexcitation initiates a coherent optical phonon mode at 4.2 THz in STO. This mode is significantly enhanced with the growth of monolayer FeSe film compared to the bare FeTe capping layers on Nb-doped SrTiO$_3$ (NSTO), providing direct evidence for the enhancement of interfacial EPC. Notably, both the relaxation process and the amplitude of the coherent optical mode observed in FeTe films exhibit anomalous changes below $T_s$ (~ 70 K), revealing a structural phase transition. Furthermore, the results of 1 UC FeSe/STO capped by FeTe suggest that STO substrate helps to stabilize FeSe structure and probably prevent the occurrence of nematic phase transition. The EPC strength λ has been estimated to be ~0.77 which is much larger than that of bulk FeSe. These results provide direct evidence for the key role of interfacial EPC in realizing the high-temperature superconductivity.

**Results**

**Quasiparticle dynamics modulation with coherent phonon dispersions.** *Ex-situ* ultrafast spectroscopy measurements were conducted on the high-quality 1 UC FeSe films grown on NSTO under protection of 15 UC FeTe films (refer to Methods for details on sample preparation), referred as 15FeTe/1FeSe/NSTO hereafter. The *in-situ* scanning tunneling microscopy (STM) characterizations of 1 UC FeSe films, before capping FeTe layers, show full gap features with high spatial homogeneity (Supplementary Fig. S1). Figure 1a shows the ultrafast pump-probe geometry with sample illustration (See more details in Methods and Supplementary Figs. S1 and S2). A white-light supercontinuum (WLC) ranging from 495 to 730 nm (1.7 - 2.5 eV) was used as probe light to monitor the electronic dynamics of intraband Fe *d* and interband Fe *d* to hybridized Fe-*d*/Se-*p* transitions, giving rise to a static optical peak around 1.8 eV[36]. Transient differential reflectivity Δ*R/R* was collected by a spectrometer at a variable delay time Δ*t*. A NIR pump light at 1100 nm (1.13 eV) was chosen to stimulate Raman phonons and avoid detection of the scattered pump leakage noise.

Upon photoexcitation of the 15FeTe/1FeSe/NSTO at 4.2 K and excitation fluence F = 2.89 mJ/cm$^2$, the two-dimensional Δ*R/R* image is modulated periodically with delay time, displaying a slow oscillation in hundreds of ps and a fast in the initial few ps (Fig. 1b). Quantitative analysis is performed by taking cuts at a probe photon energy of 1.7 eV in time windows of 0−200 ps (Fig. 1c) and 0−3 ps (Fig. 1d). By removing the incoherent electronic relaxations, the pure coherent oscillations are isolated and well fitted by exponentially decaying cosine fits with the frequencies 0.05 and 4.2 THz (Figs. 1c,1d, bottom panel), consisting with the peaks in frequency-dependent curves through fast Fourier transform (FFT) (insets to Figs. 1c,1d). We first focus on the low-frequency oscillatory response, which is a coherent acoustic phonon caused by the interference between the probe light reflected from the strain pulse propagating along the NSTO (001) direction and that reflected from the film layer, with the frequency of the oscillations depending on the phase difference between them[27,37]. Figure 1e shows the two-dimensional FFT of the oscillatory response isolated from Fig. 1b, yielding the expected linear dispersion relation of the coherent acoustic phonon in STO[37]. As clarified by previous research, this longitudinal coherent acoustic phonon provides an additional decay channel to the gluing bosons[27]. As for the high frequency 4.2 THz coherent phonon, its frequency is independent of probe photon energy (Fig. 1f) pointing to an optical phonon nature. Both electronic dynamics and FFT responses show a strong spectral weight around 1.8 eV, and quickly diminishes or even vanishes at energies around 2.2 eV, indicating a modulation of intraband Fe $d$ and interband Fe $d$ to hybridized Fe-$d$ /Se-$p$ transitions. Therefore, one reasonable conclusion to be drawn here is that a strong EPC occurs between electrons in FeTe/FeSe films and this 4.2 THz coherent optical phonon.

**Evidence for enhanced coupling between STO coherent phonon and FeSe electrons.** Considering that the penetration depth of light can be a few micrometers, and the fact that the thickness of 1UC FeTe is only 0.63 nm and that of 1UC FeSe is only 0.55 nm, the FeTe, FeSe, and NSTO components will be simultaneously excited and probed by both pump pulses and probe pulses. This naturally raises questions about the origin of the 4.2 THz optical phonon mode and how it interacts with electrons in these materials. We therefore perform control experiments on both 15 unit-cell FeTe films grown on NSTO (001) (this sample is referred as 15FeTe/NSTO hereafter) and NSTO (001) (referred as NSTO) to determine the origin of the ~ 4.2 THz mode. Figures 2a-2f show the probe phonon energy and delay time dependence of $\Delta R/R$ (Figs. 2a-2c) and the corresponding two-dimensional FFT images (Figs. 2d-2f) in 15FeTe/1FeSe/NSTO, 15FeTe/NSTO and NSTO, respectively. Under the identical incident pump excitations, a universal pump-probe interference signal in the light pulse duration near time zero occurs in all samples. The electronic relaxations accompanied by an apparent coherent oscillation at 4.2 THz are detected in both 15FeTe/1FeSe/NSTO, and 15FeTe/NSTO films but not in bare NSTO, accounting for the fact that the band gap energy of STO (~ 3.1 eV) is too large to be photoexcited by the 1100 nm (~ 1.13 eV) pump pulse. Figures 2g, 2h show that the coherent oscillation amplitude in 15FeTe/1FeSe/NSTO is more than twice as large as that in 15FeTe/NSTO, whereas the frequency does not noticeably change with photoexcited components, suggesting a much stronger coupling of this mode to the FeSe electrons than FeTe electrons. From the above discussions, we can also confirm that the coherent optical mode does not originate from the FeSe layer, considering that it also appears in the 15FeTe/NSTO film where the FeSe layer is absent. It is worth mentioning that these results are robustly reproduced by a 525 nm pump pulse (hence, still less than 3.1 eV) (Supplementary Figs. S3-S5 and Table S1).

To further determine whether the coherent optical phonon originates from FeTe or NSTO, Raman spectroscopy was employed to fingerprint the phonon vibrational modes (Fig. 2i and Supplementary Fig. S6). We only observed the Raman modes from the NSTO substrate, since the FeTe layers and FeSe monolayer are too thin to generate sufficient signals to be detected by Raman scattering. Moreover, the Raman intensity of the NSTO substrate is much suppressed by the growth of the films. STO possesses a centrosymmetric cubic perovskite structure in *Pm3m* space group above 110 K. Its Raman spectra are dominated by the second-order (2$^{nd}$) scattering, exhibiting several broad bands around 300 and 650 cm$^{-1}$ (Fig. 2i)[38-41]. Upon cooling below 110 K, STO undergoes the cubic-tetragonal antiferrodistortive (AFD) structural transition with a space group *I*4/*mcm*, leading to the appearance of first-order (1$^{st}$) modes. The AFD phase transition originates from the antiphase tilting of the Ti-O$_6$ octahedra, resulting in three new modes: one soft mode $A_{1g}$ (50 cm$^{-1}$) and two hard modes $E_g$ (144 cm$^{-1}$), $B_{1g}$ (450 cm$^{-1}$), respectively[39-43]. They are due to a folding of BZ resulting from cell doubling, whereby the phonon modes at the zone boundary R appear at the BZ center in the tetragonal phase and thus are detected by Raman spectroscopy. Noting that an additional soft mode arising near 35 cm$^{-1}$ ($E_u$) is known as ferroelectric soft mode originating from the Nb dopants microregions[26,39-43]. Compared the coherent optical phonon mode with the Raman spectroscopy results, we find that the energy of the $E_g$ (144 cm$^{-1}$) mode apparently approaches to the ~ 4.2 THz coherent optical phonon mode. As a matter of fact, the energies of the FeTe modes[44,45] are all far from 4.2 THz. We can reasonably attribute the 4.2 THz coherent optical mode to the NSTO substrate, which is associated with the antiphase tilting of the Ti-O$_6$ octahedra. What's more, this phonon can penetrate into films grown on NSTO substrate and interact with electrons therein. Generally, the $\Delta R/R$ signal of 15FeTe/1FeSe/NSTO should only slightly differ from that of 15FeTe/NSTO, since the thickness of 1 UC FeSe film is more than an order of magnitude thinner than 15 UC FeTe. However, the reflected signal of 15FeTe/1FeSe/NSTO is much stronger than that of 15FeTe/NSTO, *i.e.,* the signal of 1 UC FeSe is more dominant. On the other hand, the coherent optical mode amplitude is prominently stronger in the presence of monolayer FeSe film than in FeTe-capped NSTO substrate (see Figs. 2a-2h, Supplementary Figs. S3-S5 and Table S1), suggesting that this coherent phonon is greatly enhanced by the FeSe monolayer, which signifies direct evidence that the STO phonon couples to FeSe electrons with enhancement at the interface. It's worth noting that this coherent optical phonon mode is distinguished from the earlier high-energy F-K modes that related to the forward scattering model[4,7,25]. However, the $E_g$ mode is in-plane vibrational mode originating from the antiphase tilting of the Ti-O$_6$ octahedra. Our results provide new insights on the electron-phonon coupling mechanisim.

**Pump fluence-dependent quasiparticle dynamics.** Figure 3 presents pump fluence-dependent electronic dynamics of 15FeTe/1FeSe/NSTO, with the two-dimensional $\Delta R/R$ contour plots (Fig. 3a) versus the probe photon energy and time delays, and the corresponding FFT images (Fig. 3b), for four representative pump fluences. Under different photoexcited fluences, periodic modulations of the electronic dynamics can be observed at all pump fluences. While the dynamics evolution on pump fluence can be divided into three different regions from 0.25 to 6.3 mJ/cm$^2$. When the pump fluence is below 0.45 mJ/cm$^2$ (region I), the quasiparticle dynamics at low probe photon energy (taken a line cut at 1.7 eV shown in Figs. 3c,3d as an example), show a conspicuous periodic phonon oscillation superimposed on a positive electronic response. In this region, the stimulated Raman dominates the ultrafast dynamics. As the pump fluence increases up to intermediate 0.76-4.13 mJ/cm$^2$ (region II), a new electronic decay with a negative response emerges at $\Delta t < 1$ ps, resulting from hot carrier relaxation nonlinearly excited by pump pulses. When the pump fluence increases to the threshold above 4.13 mJ/cm$^2$ (region III), the initial positive response switches off and is replaced by a negative response. Simultaneously, a new mode located at ~3.3 THz emerges, and the amplitude of the primitive ~4.2 THz mode decreases substantially. The film has been totally destroyed at such high pump fluence.

We have extracted the dependence of the FFT amplitude and $\Delta R/R$ intensity at 6.0 ps on the pump fluence in Fig. 3e. The $\Delta R/R$ at 6.0 ps deviates from linear behavior and displays saturation behavior in region II, signifying that the measurement condition of 2.89 mJ/cm$^2$ has exceeded the low-fluence regime accompanied with thermal effects and the near-linear evolution of the temperature-dependent mode amplitude cannot represent the intrinsic ground state (Supplementary Fig. S7). The photoinduced excitations all originate from fully gapped excitations only if the $\Delta R/R$ is proportional to pump fluence[46]. Below, we restrict our measurement at 0.25 mJ/cm$^2$, which is clearly in the linear range of Fig. 3e. Significantly, the pump fluence also modulates the amplitude of coherent optical phonon with an almost concurrent change as $\Delta R/R$, as shown in Fig. 3e. That is, the coherent optical mode closely related to electronic response, further corroborating the strong EPC between substrate and monolayer FeSe.

**Temperature dependence of quasiparticle dynamics.** To further investigate the interfacial EPC between the electrons in FeSe films and phonons in the NSTO substrate, temperature-dependent pump-probe spectroscopy measurements were carried out for both 15FeTe/1FeSe/NSTO and 15FeTe/NSTO films at a low pump fluence of 0.25 mJ/cm$^2$ shown as Fig. 4. For the 15FeTe/1FeSe/NSTO sample, the contour plot of $\Delta R/R$ shown in Fig. 4a suggest that the incoherent electronic relaxations at low probe photon energy (taken 1.7 eV in Fig. 4c as an example) only consist of a positive response below ~100 K. Above 100 K, the incoherent electronic relaxation components start to exhibit a negative response in a very short time scale, while the amplitude of this negative response increases with temperature increasing. This negative response may be related to the AFD transition of STO, considering that the same feature has also been detected in 15FeTe/NSTO (Fig. 4b). Such temperature evolution can be clearly demonstrated in a line cut of the $\Delta R/R$ decay at 1.7 eV as displayed in Fig. 4c, which exhibits analogous trend to the pump fluence dependence in region II of Fig. 3c, demonstrating that the actual local temperature is higher than the nominal temperature when the pump fluence exceeds 0.45 mJ/cm$^2$ due to the thermal effect. Presumably, the actual local temperature has exceeded 200 K well above $T_c$ of monolayer FeSe films when the sample is excited at 2.89 mJ/cm$^2$ (Fig. 1 and Fig. 2). The local temperature increasing at high pump fluence can also be demonstrated by pump fluence-dependent quasiparticle dynamics in 15FeTe/NSTO (Supplementary Fig. S8). In contrast, the scenario of 15FeTe/NSTO is quite different. At 6.0 K, a negative electronic response is superimposed by a weak coherent oscillation, as displayed in Fig. 4d. As the temperature increases, the coherent oscillations gradually enhance. Moreover, $\Delta R/R$ gradually redshifts, and the amplitude of the negative response at 1.7 eV diminishes until it shuts off at ~40 K, and then turns into a positive response as the temperature increases. As the temperature further increases, the amplitude of this positive response gains strength. After reaching 100 K, the evolutionary trend becomes similar to that of FeSe.

To illustrate the evolutions of coherent oscillations that superimpose on the electronic decay, Fig. 4e summaries the temperature dependence of the amplitude of coherent optical phonon obtained from both FFT (denotes as $A_{FFT}$, right axis) and fits with an exponentially decaying cosine function (denotes $A_{cosfit}$, left axis). The open circles and open triangle data are extracted from 15FeTe/NSTO. The amplitude of the coherent optical phonon increases at first and then drops as temperature increases, resulting in an abnormal transition around 70 K. Moreover, the incoherent electronic relaxation process of the film (Supplementary Fig. S9) is synchronized with the trend of the temperature-dependent coherent phonon amplitude of the substrate, suggesting a coupling effect between electrons in the films and substrate phonons. In accordance with previous experimental results on FeTe, we assign the anomaly downtrend at ~70 K ($T_s$) to the phase transition from tetragonal (P4/*nmm*, antiferromagnetic, AFM) to monoclinic (P2$_1$/*m*, paramagnetic, PM) resulting from the antiferromagnetic order[47], and that the structural transition may lead to the change in electronic band structure and/or the variation in the scattering rate of charge carriers, thereby behaving a blue shift of Δ*R/R* below $T_S$ (Fig. 4b). It has been reported that superconducting *β*-FeSe undergoes a structure phase transition from tetragonal (P4/*nmm*, AFM)[48] to orthorhombic (*Cmma*, non-magnetic)[49] at $T_s$ ~ 90 K which originates from the electronic nematic[50,51]. It is noticeable that the quasiparticle dynamics evolutions with temperature in these two films are quite different. In the presence of monolayer FeSe, the intensity of Δ*R/R* and coherent oscillations all gradually increase with temperature decreasing, which is remarkably different from the anomaly transition in capping FeTe films grown on NSTO. The above results give compelling evidence that STO substrate is beneficial to stabilize the FeSe structure and likely prevent the occurrence of nematic phase transition, which supports that STO substrate can strongly couple to epitaxial monolayer FeSe and well modify its electronic structure, thereby realizing interfacial superconductivity enhancement.

**Discussion**

Significantly, we can directly estimate the electron-optical phonon coupling strength λ by measuring the quasiparticle lifetime. In the Allen model[52], the quasiparticle relaxation dynamics is governed by energy transfer from electrons to phonons with λ:

$$\lambda\langle\omega^2\rangle = \frac{\pi k_B T_e}{3\hbar\tau_{e-ph}}$$

where $\lambda\langle\omega^2\rangle$ is the second moment of the Eliashberg function, $\tau_{e-ph}$ is the lifetime of electron-phonon interaction that can be directly determined by ultrafast dynamics measurements, and $T_e$ is the electron temperature that can be estimated from incident pump fluence. The earlier ultrafast dynamics experiment in 1 UC FeSe/STO has obtained λ = 0.48[27]. Here, taking our quasiparticle dynamics measured coherent optical phonon frequency 4.2 THz (~17.4 meV) to replace earlier used 22 meV, λ = 0.77 can be roughly determined, which is much larger than bulk FeSe of λ = 0.16[36]. Such a large λ further corroborates that the strong interfacial EPC plays a crucial role on the interfacial high $T_c$ in the monolayer FeSe system.

In conclusion, we have observed a coherent optical phonon, corresponding to the antiphase tilting of the Ti-O$_6$ octahedra of STO substrate, in the monolayer FeSe films system. More importantly, this coherent optical mode is significantly enhanced in the presence of monolayer FeSe films, suggesting an enhanced EPC between monolayer FeSe and STO interface. A large EPC strength $\lambda = 0.77$ further reveals that interfacial EPC is responsible for the interfacial superconductivity enhancement in monolayer FeSe films. On the other hand, our pump fluence-dependent and temperature-dependent measurements indicate that STO substrate is beneficial to stabilize the FeSe structure and likely prevent the occurrence of nematic phase transition. These results support that STO substrate greatly interacts with monolayer FeSe, resulting in a strong EPC and then realizing significant interfacial superconductivity enhancement.

## Methods

**Sample preparation and STM measurement.** SrTiO$_3$ (001) substrates with 0.5% Nb doping were cleaned by direct heating at 1100 °C in an ultrahigh vacuum. Single-layer FeSe films were grown by co-evaporating high-purity Fe (99.995%) and Se (99.999%) onto substrates held at 430 °C, with a deposition rate of 12min/UC. The films were post-annealed at 470−490 °C for 3 h after growth and then transferred to the STM chamber *in-situ*. All STS measurements were carried out at 4.2 K. For *ex-situ* Raman and ultrafast spectroscopy measurements, around 15-unit-cell FeTe films were deposited on monolayer FeSe grown on Nb-doped SrTiO$_3$ (001) by co-evaporation of Fe and Te onto the substrate kept at ~250 °C, with a deposition rate of 12min/UC. In addition, 15 UC FeTe grown on Nb-doped SrTiO$_3$ (001) films were prepared for a parallel control experiment.

**Raman spectroscopy characterization.** Raman spectroscopy measurements were carried out by using a Horiba iHR550 spectrometer equipped with a liquid-nitrogen cooled CCD. A 532 nm laser was used as excitation. The low-frequency Raman spectrum below 50 cm$^{-1}$ was collected by ultralow wave number mode. The laser power was maintained at ~1.0 mW in all measurements.

**Ultrafast spectroscopy measurement.** 800 nm, 35 fs femtosecond pulses were generated by a femtosecond laser at a repetition rate of 1 kHz (Coherent Ti sapphire Astrella amplifier system). The pulses were split into pump and probe beams. The pump beam entered the optical parametric amplifier (OPA, Coherent opera solo system) to generate a wide range of adjustable pump pulses (240 nm−15 μm) providing conditions for various pump excitations. The probe beam passed through a delay line to vary pump-probe delay time, and was further focused into a nonlinear crystal (sapphire) to generate white light supercontinuum (WLC, 420−980 nm), realizing the probing of time-resolved reflection spectroscopy in the visible range. The reflected probe light is collected by a high-performance spectrometer running at 1 kHz. The overall time resolution was about 150 fs. Equipped with vacuum and low-temperature systems, variable temperature measurements from 4.2 to 325 K can be realized. Both 1100 nm and 525 nm pump pulses were used for intraband and interband excitations. After pump excitation, the transient differential reflectivity $\Delta R/R$ was recorded as a function of the delay time of the probe pulses and probe phonon energy. The pump fluence was kept at a range of 0.25−6.3 mJ/cm$^2$ for pump fluence-dependent measurements.

## Data availability

The data that support the findings of this paper are available from the corresponding authors upon request.

**Acknowledgements**

This work was supported the National Natural Science Foundation of China (Grant Nos. 12020101003, 12250710126, 92056204, 52388201), and the National Basic Research Program of China (Grant No. 2021YFA1400100, 2022YFA1403102).

**Author contributions**

⊥ M.Z., and Q.X. proposed and conceived this research. M.Z. performed Raman measurements and ultrafast spectroscopy experiments, developed ultrafast spectroscopy system, analyzed data, and wrote the paper. X.J., M.S., W.D., C.D. and L.W. performed thin-film growth and STM measurements. Y.W. performed Raman measurements. H.L. analyzed data. All authors participated in the discussion of results and commented on the paper.

**Competing interests**

The authors declare no competing interest.

**Additional information**

Correspondence and requests for materials should be addressed to Qihua Xiong (qihua_xiong@mail.tsinghua.edu.cn).

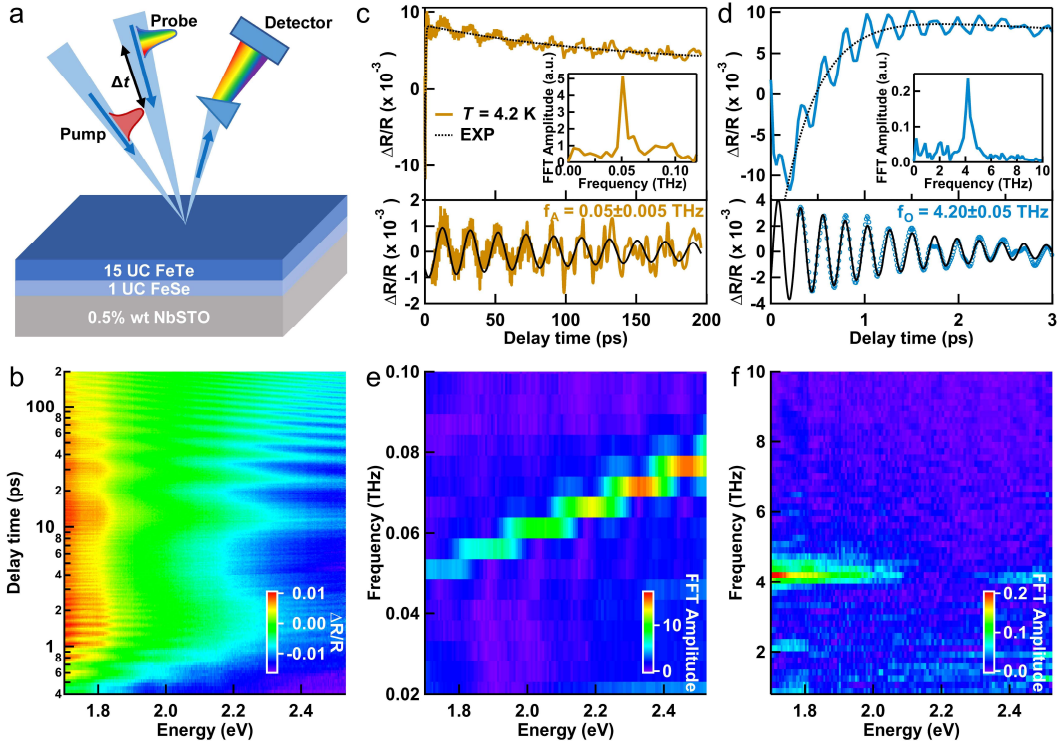

**Fig. 1. Experimental geometry and quasiparticle dynamics modulation with coherent phonon dispersions. a**, Monolayer FeSe grown on NSTO (001) protected by 15-unit-cels FeTe capping layers was measured by ultrafast pump-probe spectroscopy. $\Delta t$ denotes the delay of the WLC probe pulse with respect to the NIR pump pulse. **b**, Two-dimensional contour plot of transient differential reflectivity $\Delta R/R$ as a function of delay $\Delta t$ and probe photon energy at 4.2 K. The incident pump wavelength for all measurements is 1100 nm unless stated otherwise. The pump fluence is 2.86 mJ/cm$^2$. The delay time axis is plotted on a logarithmic scale. **c, d**, Cuts of two-dimensional $\Delta R/R$ in **b** showing $\Delta R/R$ dynamics at a selected probe photon energy of 1.7 eV in time windows of 0-200 ps (**c**) and 0-3 ps (**d**). Time-resolved coherent phonons were superimposed on electronic relaxation. Dashed black curves are single (**c**) and double exponential fits (**d**) of the incoherent electronic relaxations, respectively. The superimposed coherent oscillations after subtracting the electronic relaxations are further displayed in the bottom panel. Solid black line: corresponding fits with an exponentially decaying cosine. Inset: Fast Fourier transform. **e, f**, Two-dimensional FFT images of coherent oscillations corresponding to time windows of 0-200 ps (**e**) and 0.3-3 ps (**f**), respectively.

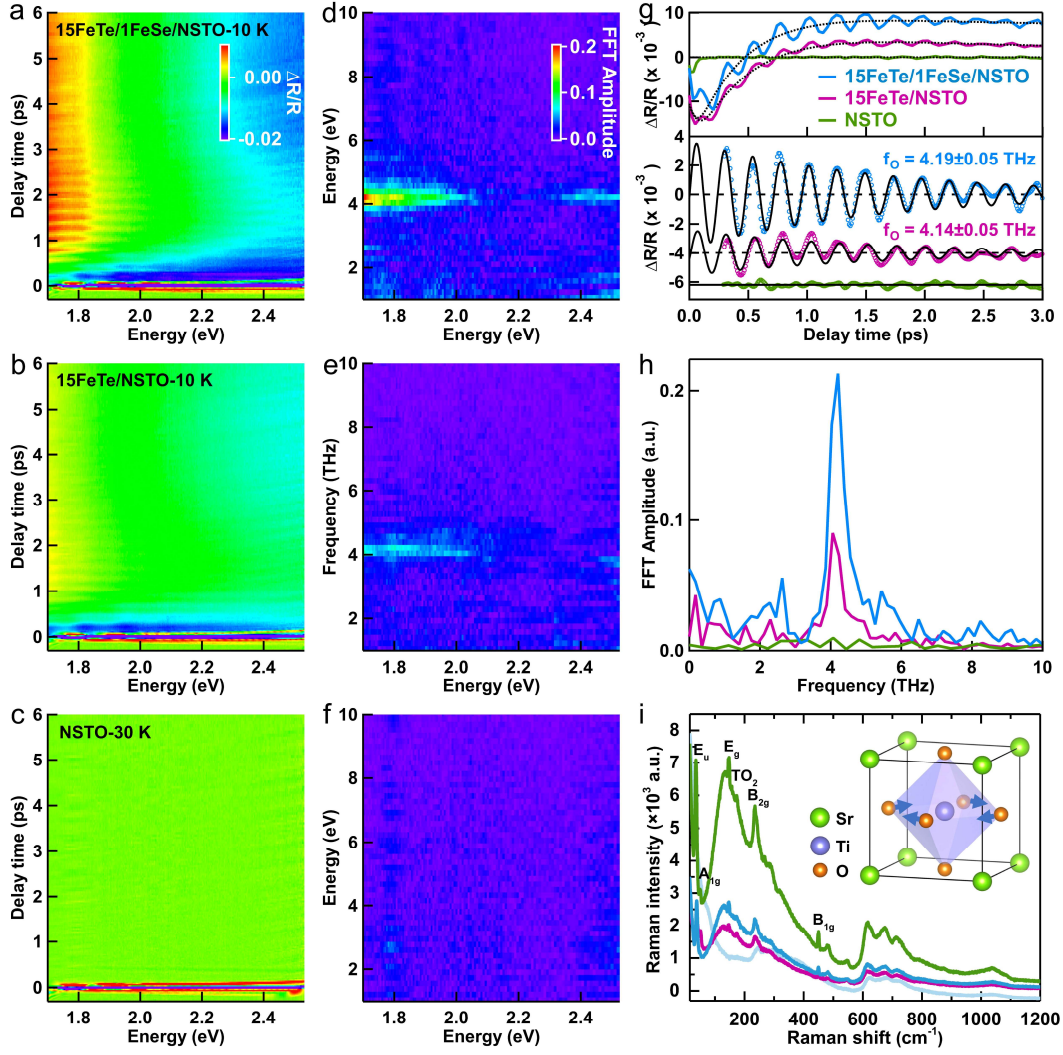

**Fig. 2. Significant enhancement of the coherent optical phonon amplitude in 15FeTe/1FeSe/NSTO compared to 15FeTe/NSTO and bare NSTO. a-c**, Two-dimensional contour plot of $\Delta R/R$ measured in 15FeTe/1FeSe/NSTO at 10 K (**a**), 15FeTe/NSTO at 10 K (**b**), and bare NSTO at 30 K (**c**), with the same pump fluence at 2.86 mJ/cm$^2$ and identical color scale. **d-f**, Two-dimensional FFT of the coherent oscillations in **a**, **b**, and **c**, after subtracting the incoherent signal. **g**, Time-dependent $\Delta R/R$ at probe photon energy 1.7 eV cut from **a**, **b**, and **c**. Dashed black curves indicate dual-exponential fits. Extracted coherent phonon oscillations are further displayed in the bottom panel with a vertical offset. Solid black lines indicate the fitting by a cosine function multiplied with an exponential envelope. **h**, Corresponding FFT curves of the pure oscillatory response in the lower panel of **g**. **i**, Comparison of Raman spectra for bare STO (green line), FeTe film (purple line) and FeSe film (blue line) at 5.5 K. The light blue line is 15FeTe/1FeSe/NSTO film at 250 K. Inset: Schematic of the ionic vibration of Raman modes induced by STO AFD structural transition. All samples were excited by a 532 nm laser with a power of 1 mW in all measurements.

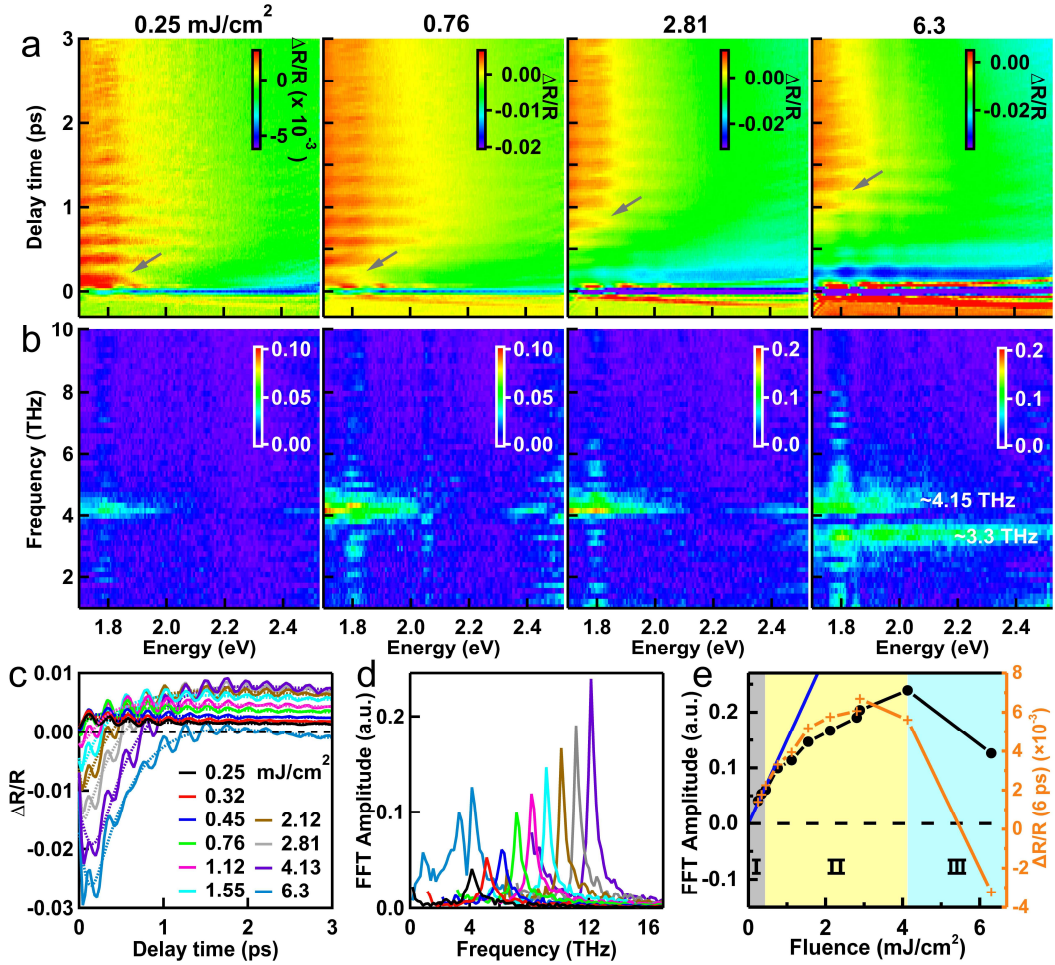

**Fig. 3. Dependence of ultrafast dynamics and coherent optical phonon dispersions on pump fluence. a**, Two-dimensional contour plot of the $\Delta R/R$ measured in 15FeTe/1FeSe/NSTO excited by four representative pump fluence at 6 K. **b**, Two-dimensional FFT images as a function of both frequency and probe photon energy, corresponding to $\Delta R/R$ images in **a**. **c**, Dependence of $\Delta R/R$ dynamics on the incident pump fluence cut at 1.7 eV probe photon energy. **d**, Dependence of FFT curves on the incident pump fluence transformed from **c**. The curves from 0.25-4.13 mJ/cm$^2$ are offset by 1 THz horizontally for clarity. **e**, FFT amplitude (left axis, black) and $\Delta R/R$ intensity at 6 ps (right axis, origin) as a function of pump fluence. Solid blue line is a linear guide to the eye.

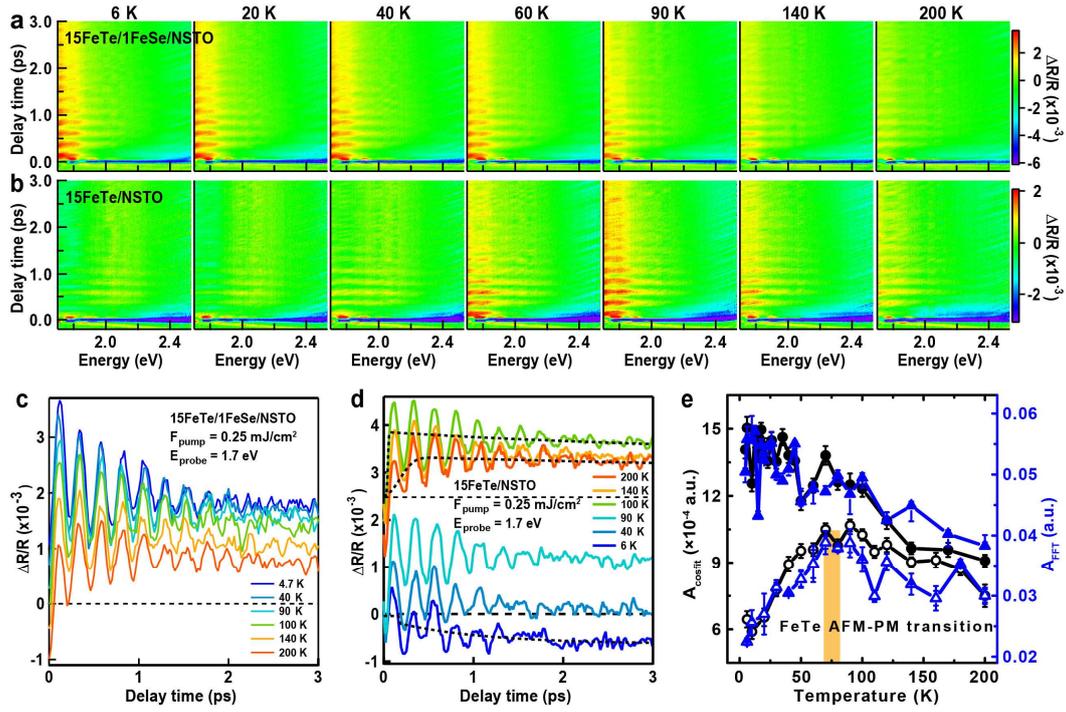

**Fig. 4. Temperature-dependent quasiparticle dynamics. a,b,** Two-dimensional contour plot of the $\Delta R/R$ images taken at pump fluence 0.25 mJ/cm$^2$ from 6 to 200 K in 15FeTe/1FeSe/NSTO (**a**) and 15FeTe/NSTO (**b**), respectively. **c,d,** Dependence of the $\Delta R/R$ dynamics on temperature cut at 1.7 eV extracted from (**a**) and (**b**), respectively. The curves from 100 K to 200 K of 15FeTe/NSTO in **d** are vertically offset by 0.0025 for clarity. **e,** Temperature-dependent amplitudes obtained by cosine fit (left axis, black) and FFT (right axis, blue) extracted from the analysis of data in **c** and **d**. The solid circles and triangles are extracted from 15FeTe/1FeSe/NSTO, and the open ones are extracted from 15FeTe/NSTO. The orange stripe around 70 K marks the AFM-PM phase transition of FeTe.

Supplementary Information for

# Direct evidence of coherent electron-phonon coupling in single-unit-cell FeSe film on Nb-doped SrTiO$_3$

Mengdi Zhang[1,2], Xiaotong Jiao[2], Mingxia Shi[2], Wenfeng Dong[2], Cui Ding[1,2], Yubin Wang[2], Tingxiao Qin[1], Haiyun Liu[1*], Lili Wang[2,3*], Zhenyu Zhang[5], Qi-Kun Xue[1,2,3,6], Qihua Xiong[1,2,3,4*]

[1] *Beijing Academy of Quantum Information Sciences, Beijing 100193, China*

[2] *State Key Laboratory of Low-Dimensional Quantum Physics and Department of Physics, Tsinghua University, Beijing 100084, China*

[3] *Frontier Science Center for Quantum Information, Beijing 100084, China*

[4] *Collaborative Innovation Center of Quantum Matter, Beijing 100084, China*

[5] *International Center for Quantum Design of Functional Materials (ICQD), Hefei National Laboratory for Physical Sciences at Microscale, and Synergetic Innovation Center of Quantum Information and Quantum Physics, University of Science and Technology of China, Hefei, Anhui 230026, China*

[6] *Southern University of Science and Technology, Shenzhen 518055, China*

# Table of Contents

**Supplementary Fig. S1**

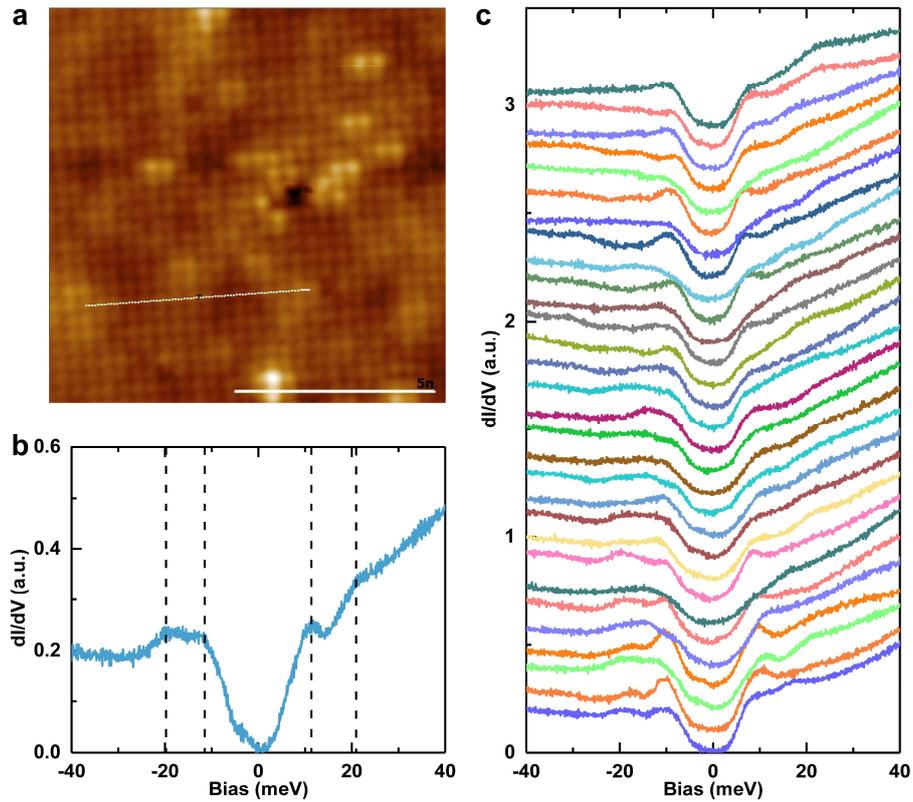

**Supplementary Fig. S1. Surface topography and superconducting gap spectra of 1 UC FeSe grown on Nb-doped SrTiO$_3$ (001) (NSTO) before capping FeTe layers. a**, An STM topography of 1FeSe/NSTO (10×10 nm$^2$, $V_b$ = 50 mV, $I$ = 500 pA). **b**, A typical tunneling spectrum taken on the film at 4.2 K. Fully gaped superconductivity accompanied by two pairs of coherent peaks are marked in dashed black lines at $\varDelta_1$ ~ 11 meV and $\varDelta_2$ ~ 20 meV. **c**, A series of spectra taken along a line as marked in **a**, confirming the spatial homogeneity of superconductivity.

**Supplementary Figure S2**

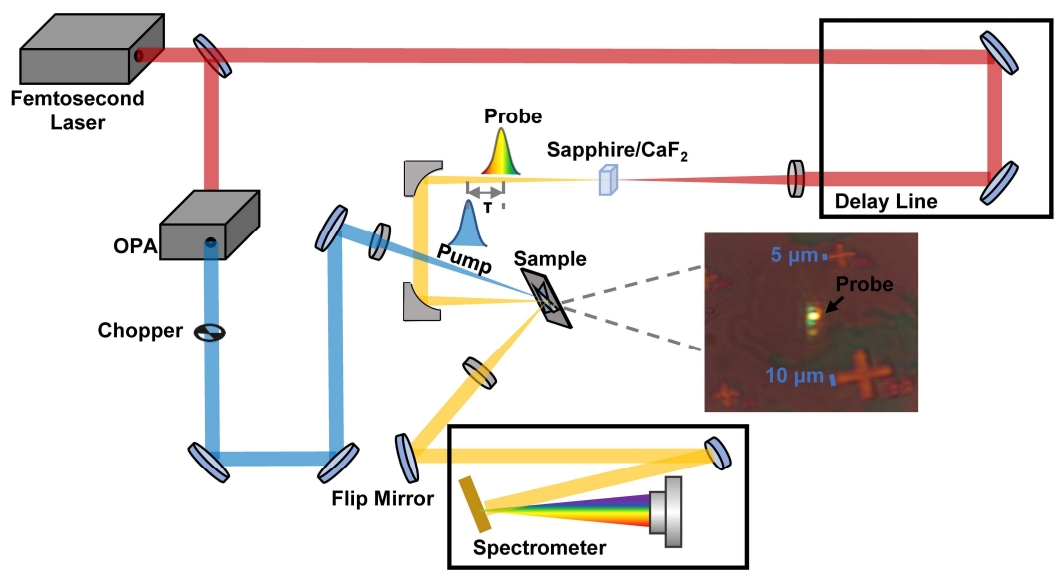

**Supplementary Fig. S2. Experimental setup for ultrafast pump-probe spectroscopy schematic.** A femtosecond laser beam (800 nm, 35 fs, 1 kHz repetition rate, Coherent Ti sapphire astrella amplifier system) is split into two parts. One part goes to an optical parameter amplifier (OPA, Coherent opera solo system) to generate pump pulses at 1100 nm. The other part goes to a delayline to vary pump-probe delay time, and is further focused into a sapphire crystal in order to produce white light continuum (WLC) for the probe. The reflected probe light is collected by a high-performance spectrometer running at 1 kHz.

**Supplementary Fig. S3**

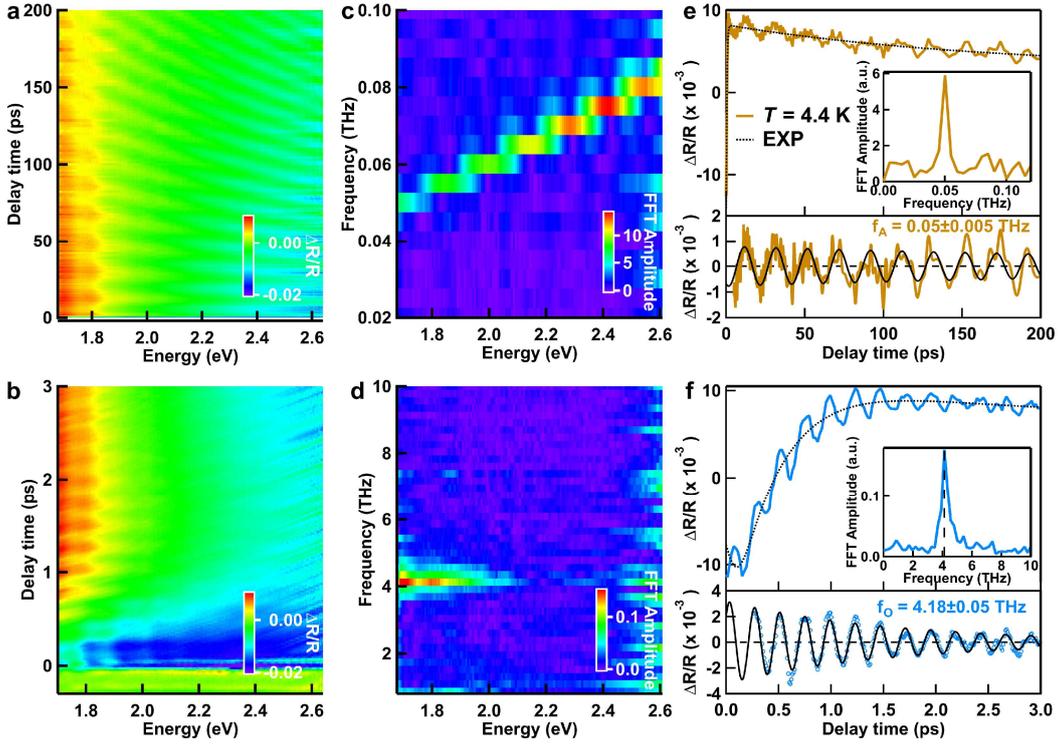

**Supplementary Fig. S3. Quasiparticle dynamics modulation with coherent phonon dispersions by 525 nm pump in 15FeTe/1FeSe/NSTO film. a,b,** Two-dimensional contour plot of $\Delta R/R$ as a function of both pump-probe delay $\Delta t$ and probe photon energy at 4.2 K in time windows of 0-200 ps (**a**) and 0-3 ps (**b**), respectively. The incident pump wavelength is 525 nm. The pump fluence is 2 mJ/cm$^2$. **c,d,** Two-dimensional FFT from images in **a** and **b** after subtracting the incoherent electronic relaxations. **e,f,** Cuts of the two-dimensional $\Delta R/R$ mapping in **a** and **b** showing $\Delta R/R$ intensity dynamics at selected probe photon energy 1.7 eV in time windows of 0-200 ps (**e**) and 0-3 ps (**f**), respectively. Time-resolved coherent phonons were superimposed on electronic relaxation. Dashed black curves correspond to single- (**e**) and dual-exponential (**f**) fittings, respectively. In the lower panels, the superimposed low-frequency (**e**) and high-frequency coherent phonon (**f**) are obtained by subtracting the electronic relaxations. Solid black line: corresponding fits with an exponentially decaying cosine. Inset: FFT curves. Upon photoexcitation by 525 nm pump pulse, the two-dimensional $\Delta R/R$ mapping is also periodically modulated with delay time, which is akin to that by 1100 nm pump, displaying two strong oscillatory components.

**Supplementary Fig. S4**

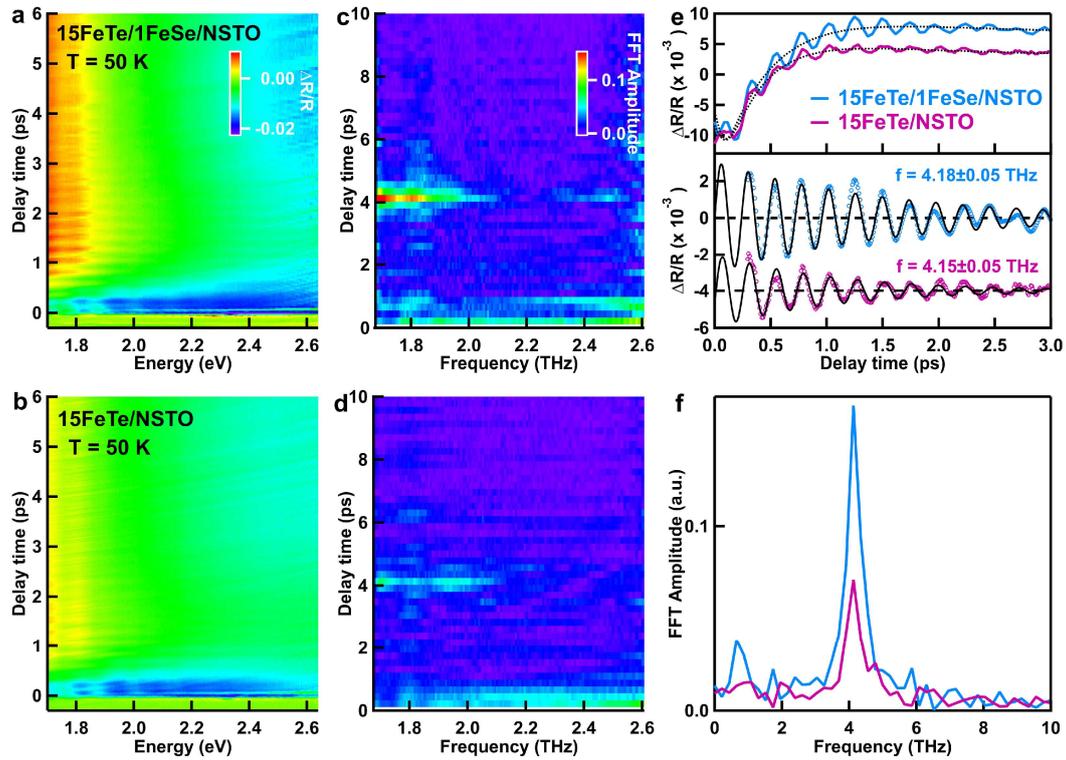

**Supplementary Fig. S4. Significant enhancement of the coherent optical phonon amplitude in 15FeTe/1FeSe/NSTO compared to 15FeTe/NSTO excited by 525 nm. a,b**, Two-dimensional contour plot of $\Delta R/R$ measured in 15FeTe/1FeSe/NSTO (**a**) and 15FeTe/NSTO (**b**) at 10 K, with the same color scale and a pump fluence at 2 mJ/cm². **c,d**, Two-dimensional FFT of the $\Delta R/R$ mapping after subtracting the incoherent signals in **a** and **b**, respectively. **e**, $\Delta R/R$ intensity dynamics at probe energy 1.7 eV cut from **a** and **b**. Dashed black curves indicate two exponential fits of the data. Extracted coherent phonon oscillations are further displayed in the lower panel. Solid black lines indicate the fitting by a cosine function multiplied with an exponential envelope, and curves are offset for clarity. **f**, Corresponding FFT curves of the pure oscillatory responses in the lower panel of **e**. Similar to photoexcitation by 1100 nm, here the amplitude of coherent optical phonon detected in 15FeTe/1FeSe/NSTO film is significantly stronger than that in 15FeTe/NSTO, giving direct evidence of that the STO phonon strongly couples to FeSe electrons at the interface, and further demonstrating the robustness of our results as well.

**Supplementary Fig. S5**

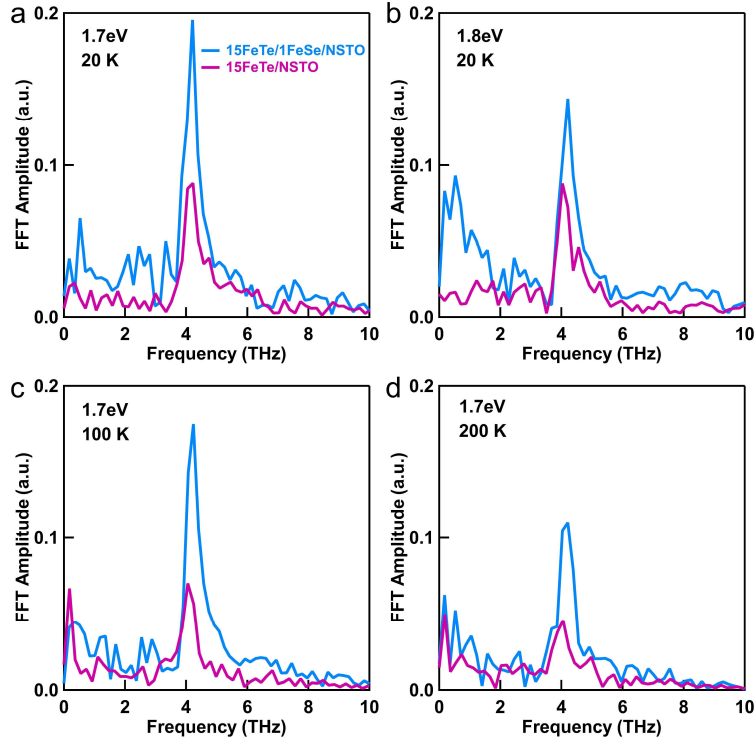

**Supplementary Fig. S5. Enhancement of coherent optical phonon amplitude in the presence of FeSe monolayer at different measure conditions. a-d**, Comparisons of the coherent optical phonon detected in 15FeTe/1FeSe/NSTO and 15FeTe/NSTO at virous photoexcited conditions. (**a**) $T$ = 20 K, probe photo energy = 1.7 eV, (**b**) $T$ = 20 K, probe photon energy = 1.8 eV, (**c**) $T$ = 100 K, probe photo energy = 1.7 eV and (**d**) $T$ = 200 K, probe photon energy is 1.7 eV. The incident pump wavelength is all at 1100 nm and the pump fluences are 2.89 mJ/cm$^2$.

**Table S1. Comparisons of the coherent optical phonon detected in 15FeTe/1FeSe/NSTO and 15FeTe/NSTO at virous photoexcited conditions.** Here $\lambda_{pump}$, $E_{probe}$, $f_o$ and $A_{FFT}$ denote the pump wavelength, the probe photon energy, the frequency of coherent optical phonon, and the amplitude of coherent optical phonon, respectively.

| $\lambda_{pump}$ (nm) | | 1100 | | | | | 525 | |
|---|---|---|---|---|---|---|---|---|
| $T$ (K) | | 6 | 20 | 20 | 100 | 200 | 50 | 50 |
| $E_{probe}$ (eV) | | 1.7 | 1.7 | 1.8 | 1.7 | 1.7 | 1.7 | 1.8 |
| 15FeTe/1FeSe/NSTO | $f_o$ (THz) | 4.19 | 4.19 | 4.19 | 4.19 | 4.19 | 4.18 | 4.18 |
| | $A_{FFT}$ | 0.1926 | 0.1954 | 0.1435 | 0.1749 | 0.1097 | 0.1659 | 0.1369 |
| 15FeTe/NSTO | $f_o$ (THz) | 4.18 | 4.18 | 4.18 | 4.18 | 4.18 | 4.15 | 4.15 |
| | $A_{FFT}$ | 0.0876 | 0.0929 | 0.0878 | 0.0678 | 0.0562 | 0.0716 | 0.0436 |

**Supplementary Fig. S6**

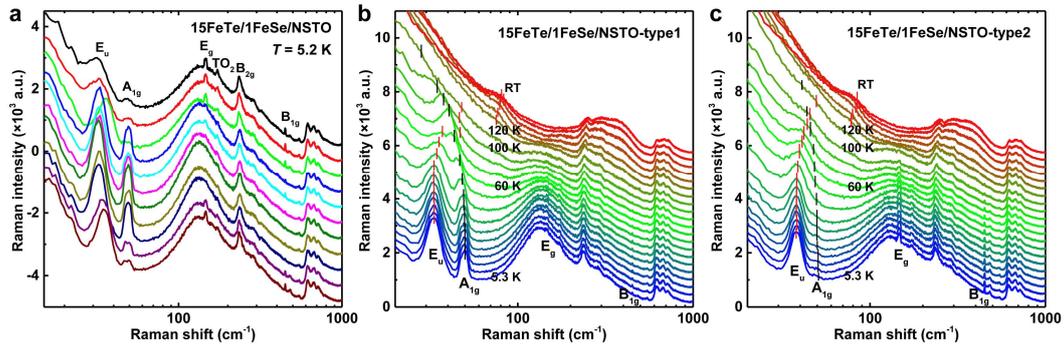

**Supplementary Fig. S6. Region-dependent and temperature-dependent Raman spectra of 15FeTe/1FeSe/NSTO films. a,** A series of Raman spectra detected in different regions at 5.2 K. In different regions, the intensity of soft mode $A_{1g}$ and hard modes $E_g$ and $B_{1g}$ behave quite differently. It's not difficult to spot that the intensity of soft mode $A_{1g}$ is in inversely proportion to the intensity of hard modes $E_g$ and $B_{1g}$, producing two representative types of Raman spectra. Type1: the intensity of $A_{1g}$ mode is relatively high, while the intensity of $E_g$ and $B_{1g}$ modes approach to vanish. Type2: the intensity of $A_{1g}$ mode is faint, but the intensity of $E_g$ and $B_{1g}$ modes are evident. **b,c,** Temperature evaluations of type1 (**b**) and type2 spectra (**c**) from 5.3 K to room temperature (RT). On warming, hard modes $E_g$ and $B_{1g}$ decrease in intensity and vanish above ~ 110 K (AFD transition temperature), soft mode $A_{1g}$ softens and decreases in intensity until fade away above 110 K, ferroelectric soft mode $E_u$ exhibits a quite different behaviors, which hardens and undergoes an anomaly change in energy at around 60 K, leading to a ferroelectric transition.

**Supplementary Fig. S7**

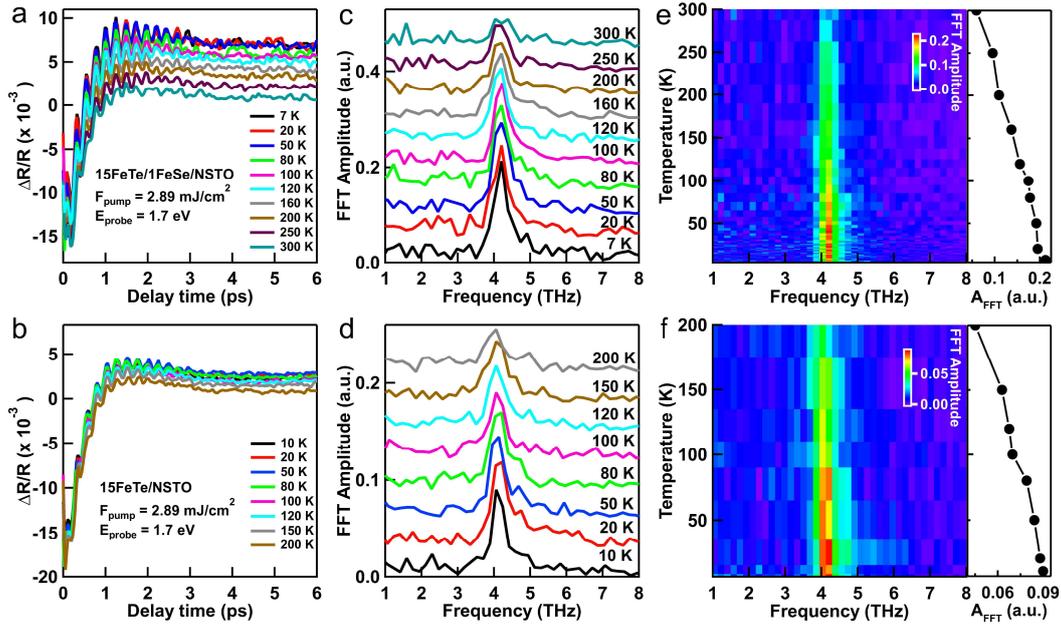

**Supplementary Fig. S7. Temperature evolutions of quasiparticle dynamic in 15FeTe/1FeSe/NSTO and 15FeTe/NSTO photoexcited by 1100 nm pump with high pump fluence 2.89 mJ/cm$^2$. a,b**, Dependence of $\Delta R/R$ dynamics on temperature at probe phonon energy 1.7 eV in 15FeTe/1FeSe/NSTO and 15FeTe/NSTO, respectively. **c,d**, FFT curves at different temperatures obtained from **a** and **b**, respectively. **e,f**, Two-dimensional FFT amplitudes as a function of both temperature and FFT frequency corresponding to 15FeTe/1FeSe/NSTO and 15FeTe/NSTO, respectively. The right panel denotes the relation of FFT amplitude on temperature at probe phonon energy 1.7 eV. For two films, the absolute intensity of ΔR/R and coherent oscillations all gently decrease with temperature increasing. It is necessary to note that above measurements were performed under a relatively high excitation pump fluence of 2.86 mJ/cm$^2$, which have already induced thermal effects in the materials. That is, temperature-dependent quasiparticle dynamic at 2.89 mJ/cm$^2$ cannot represent the intrinsic ground state.

**Supplementary Fig. S8**

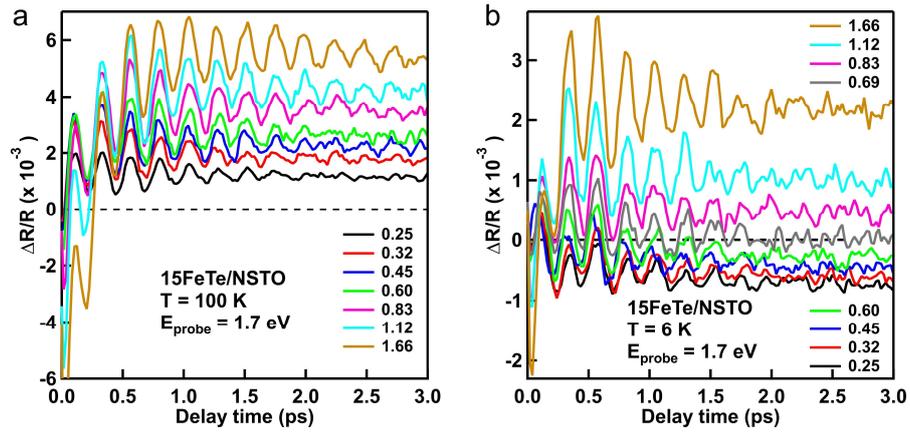

**Supplementary Fig. S8. Dependence of quasiparticle relaxations on incident pump fluence below and above phase transition temperature in 15FeTe/NSTO. a,b**, Dependence of $\Delta R/R$ dynamics on incident pump fluence cut at probe photon energy 1.7 eV at 100 K (**a**) and 6 K (**b**), respectively. The pump fluence changes from 0.25 to 1.66 mJ/cm$^2$.

**Supplementary Fig. S9**

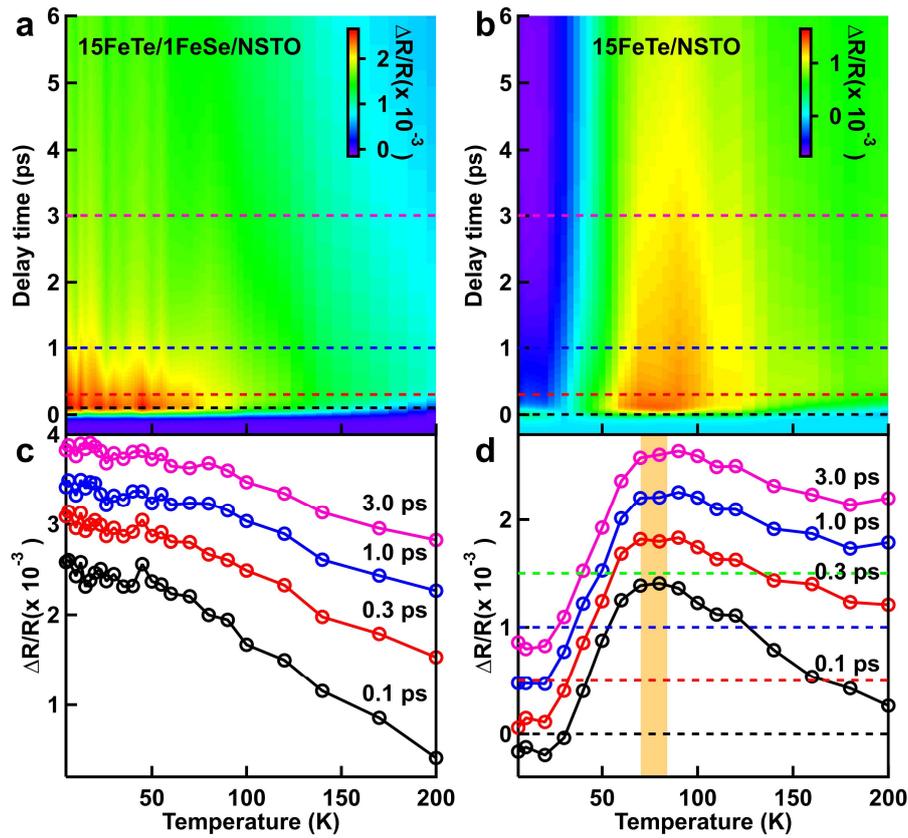

**Supplementary Fig. S9. Dependence of incoherent electronic relaxations on temperature in 15FeTe/1FeSe/NSTO and 15FeTe/NSTO photoexcited by low pump fluence. a,b**, Temperature and delay time dependence of incoherent electronic relaxations at probe phonon energy 1.7 eV in 15FeTe/1FeSe/NSTO and 15FeTe/NSTO, respectively. The incoherent electronic relaxations are isolated from the dual-exponential fittings. **c,d**, Incoherent $\Delta R/R$ as a function of temperature at various delay times of 0.1, 0.3, 1 and 3 ps extracted from **a** and **b**, respectively. The orange stripe marks the AFM-FM phase transition of FeTe, which is akin to Fig. 4e. For 15FeTe/1FeSe/NSTO, $\Delta R/R$ of incoherent electronic relaxations gradually decreases with temperature increase. In contrast, in 15FeTe/NSTO, ΔR/R of incoherent electronic relaxations increase at first and then drop down as temperature increasing, resulting in an abnormally transition around 70 K. This transition of electronic relaxations in sync with the coherent optical mode in Fig. 4e, indicating the AFM-FM phase transition in FeTe.